\documentclass[conference]{IEEEtran}

\usepackage{graphicx}
\usepackage{hyperref}
\usepackage{subfigure}
\usepackage{amsmath}
\usepackage{algorithm}
\usepackage{algorithmicx}
\usepackage{algpseudocode}
\usepackage{arydshln}

\graphicspath{{./images/}}

\title{Exploring the structure and function of temporal networks with dynamic graphlets}

\author{
\textbf{Yuriy Hulovatyy$^{1,2,3}$, Huili Chen$^{1}$, and Tijana Milenkovi\'{c}$^{1,2,3,*}$}\\
$^1$Department of Computer Science and Engineering\\
$^2$Interdisciplinary Center for Network Science and Applications
(iCeNSA)\\ $^3$ECK Institute for Global Health\\ University of Notre
Dame, Notre Dame, IN 46556, USA\\ $^*$Corresponding Author (Email:
tmilenko@nd.edu) }

\begin{document}
\maketitle

\begin{abstract}
With the growing amount of available temporal real-world network data,
an important question is how to efficiently study these data. One can
simply model a temporal network as either a single aggregate static
network, or as a series of time-specific snapshots, each of which is
an aggregate static network over the corresponding time window. The
advantage of modeling the temporal data in these two ways is that one
can use existing well established methods for static network analysis
to study the resulting aggregate network(s). However, doing so loses
valuable temporal information either completely (in the first case) or
at the interface between the different snapshots (in the second case).
Here, we develop a novel approach for studying temporal network data
more explicitly, by using the snapshot-based representation but by also
directly capturing relationships between the different snapshots. We
base our methodology on the well established notion of graphlets
(subgraphs), which have been successfully used in numerous contexts in
static network research. Here, we take the notion of static graphlets
to the next level and develop new theory needed to allow for
graphlet-based analysis of temporal networks. Our new notion of
dynamic graphlets is quite different than existing approaches for
dynamic network analysis that are based on temporal motifs
(statistically significant subgraphs). Namely, these approaches suffer
from many limitations. For example, they can only deal with subgraph
structures of limited complexity. Also, their major drawback is that
their results heavily depend on the choice of a null network model
that is required to evaluate the significance of a subgraph. However,
choosing an appropriate null network model is a non-trivial task. Our
dynamic graphlet approach overcomes the limitations of the existing
temporal motif-based approaches. At the same time, when we thoroughly
evaluate the ability of our new approach to characterize the structure
and function of an entire temporal network or of individual nodes, we
find that the dynamic graphlet approach outperforms the static
graphlet approach, independent on whether static graphlets are applied
to the aggregate network or to the snapshot-based network
representation. Clearly, accounting for more temporal information
helps with result accuracy.

\end{abstract}

\section{Introduction}
\subsection{Motivation}
Networks (or graphs) are powerful models for studying complex systems
in various domains, from biological cells to societies to the
Internet.  Traditionally, due to limitations of data collection
techniques, researchers have mostly focused on studying the static
network representation of a given system
\cite{newman2010networks}.
However, many real-world systems are not static but they change over
time
\cite{holme2012temporal}.  With new technological advancements, it has
become possible to record temporal changes in network structure (or
topology).  Thus, on top of the traditional static network
representation of the system of interest, one can now also obtain
information on arrival or departure times of nodes or edges. Examples
of temporal networks include person-to-person communication
\cite{priebe2005scan}, online social 
\cite{leskovec2008microscopic}, citation 
\cite{leskovec2005graphs}, cellular 
\cite{przytycka2010toward}, and functional brain 
\cite{valencia2008dynamic} networks.

The increasing availability of temporal real-world network data, while
opening new opportunities, has also raised new challenges for
researchers.  Namely, despite a large arsenal of powerful methods that
already exist for studying static networks, these methods cannot be
directly applied to temporal network data. Instead, the simplest
approach to deal with a temporal network is to completely discard its
time dimension by \emph{aggregating} all nodes and edges from the
temporal data into a single static network.  While this would allow to
directly apply to the resulting aggregate network the existing and
well established methods for static network analysis, such an
aggregate or
\emph{static} approach loses all important temporal information from the data.
To overcome this, one could model the temporal network as a
\emph{series of snapshots}, each of which is a static network that
aggregates the temporal data observed during the corresponding time
interval. Then, with such a snapshot-based network representation, one
could use a \emph{static-temporal} approach to study each snapshot
independently via the existing methods for static network analysis and
then consider time-series of the results. However, this strategy
treats each network snapshot in isolation and discards relationships
between the different snapshots.  Clearly, both static and
static-temporal approaches overlook temporal information that is
important for studying evolution of a dynamic system
\cite{holme2012temporal}.
Therefore, proper analysis of temporal network data requires
development of conceptually novel strategies that can fully exploit
the available temporal information from the data. And this is exactly
the focus of our study.

\subsection{Related work}\label{sec:related_work}
\begin{figure}
  \centering\includegraphics[width=\linewidth]{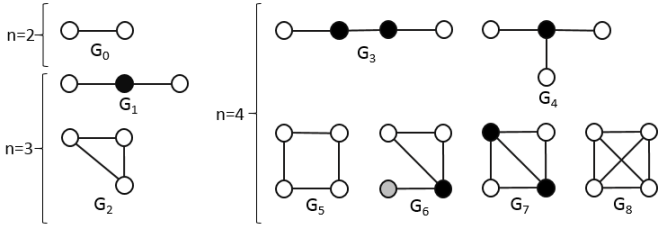}
  \caption{All nine static graphlets with up to four nodes, along with
  their 15 ``node symmetry groups'' (or formally, automorphism orbits)
  \cite{prvzulj2004modeling,milenkovic2008uncovering}. Within a given
  graphlet, different orbits are denoted by different node colors.
  For example, there is a single orbit in graphlet $G_2$, as all three
  nodes are topologically identical to each other. But there are two
  orbits in graphlet $G_2$, as the two end nodes are topologically
  identical to each other but not to the middle node (and vice versa).
  }\label{fig:static_graphlets}
\end{figure}

\noindent\textbf{Static networks.}
One way to study the structure of a static network is to compute its
\textit{global} properties such as the degree distribution, diameter,
or clustering coefficient
\cite{newman2010networks}.  However, even though global network
properties can summarize the structure of the entire network in a
computationally efficient manner, they are not sensitive enough to
capture detailed topological characteristics of the complex real-world
networks \cite{GraphCrunch}.  Thus, \textit{local}
properties have been proposed that can capture more detailed aspects
of complex network structure. For example, one can study small
\emph{partial} subgraphs called \textit{network motifs} that are
statistically significantly over-represented in a network compared to
some null model \cite{milo2002network,milo2004superfamilies}.
However, the practical usefulness of network motifs has been
questioned, since the choice of null model can significantly affect
the results
\cite{artzy2004comment,GraphCrunch}, and since selecting 
an appropriate null model is not a trivial task
\cite{milenkovic2009optimized}.
Hence, to address this challenge, \textit{graphlets} have been
proposed \cite{prvzulj2004modeling}, which are small \emph{induced}
subgraphs of a network that can be employed without reference to a
null model (\figurename~\ref{fig:static_graphlets}), unlike network
motifs.  Also, unlike network motifs, graphlets must be induced
subgraphs, whereas motifs are partial subgraphs, which makes graphlets
more precise measures of network topology compared to motifs
\cite{GraphCrunch}.  Graphlets have been well established 
when studying static networks. For example, they were used as a basis
for designing topologically constraining measures of network
\cite{prvzulj2004modeling} or node 
\cite{milenkovic2008uncovering} similarities. These measures 
in turn have been used to develop state-of-the-art algorithms for
various computational problems such as network alignment
\cite{milenkovic2010optimal,GRAAL,MAGNA}, clustering
\cite{Solava2012graphlet,hulovatyy2014network}, or de-noising
\cite{hulovatyy2014revealing}, as well as for a number of application problems, such as studying human aging \cite{Faisal2014dynamic,Faisal2014a}, cancer \cite{milenkovic2010systems,Ho2010}, or pathogenicity \cite{Solava2012graphlet,Milenkovic2011}.

\noindent\textbf{Temporal networks.}
Analogous to studying static networks from a global perspective,
temporal networks can also be studied by considering evolution of
their global properties
\cite{leskovec2005graphs,nicosia2012components}. 
As this again leads to imprecise insights into network changes with
time, recent focus has shifted onto local-level dynamic network
analysis via notion of ``temporal motifs''. In the simplest case of
the static-temporal approach, static motifs (as defined above) are
counted in each snapshot and then their counts are compared across the
snapshots
\cite{braha2009time}. To overcome this approach's  limitation of
ignoring any motif relationships between different snapshots, the
notion of static network motifs has been extended into several notions
of temporal motifs
\cite{zhao2010communication,bajardi2011dynamical,kovanen2011temporal,kovanen2013temporal,jurgens2012temporal}.  However, each
of these existing temporal motif-based approaches suffers from at
least three of the following drawbacks:

\noindent\textbf{1.} They can only deal with motif structures of limited complexity, 
such as small motifs or simple topologies (e.g., linear paths)
\cite{zhao2010communication,bajardi2011dynamical,jurgens2012temporal}, 
which limits their practical usefulness to capture complex network
structure in detail.

\noindent\textbf{2.} They pose  additional constraints, such as limiting 
the number of events (temporal edges) a node can participate in at a given time point
\cite{kovanen2011temporal,kovanen2013temporal}.

\noindent\textbf{3.}  They allow for obtaining the motif-based topological ``signature'' of 
 the entire network only but \emph{not} of each individual node
\cite{braha2009time,zhao2010communication,bajardi2011dynamical,jurgens2012temporal,kovanen2011temporal,kovanen2013temporal}, whereas the latter is
very useful when aiming to link the network topological position of a
node to its function via e.g., network alignment or clustering (see
the above discussion on static graphlets).

\noindent\textbf{4.}  Importantly, just as static motifs,  the temporal 
motif-based approaches rely on a null model
\cite{braha2009time,zhao2010communication,bajardi2011dynamical,jurgens2012temporal,kovanen2011temporal,kovanen2013temporal}, which again makes their practical usefuleness questionable \cite{artzy2004comment,GraphCrunch,milenkovic2009optimized}, especially since 
adequately choosing an appropriate null model is an even more complex
problem in the dynamic setting compared to the static setting (see
above).

\noindent\textbf{5.}  They are typically dealing with directed networks 
(e.g., mobile communications). While this is not necessarily a
drawback \emph{per se}, many real-world networks are
\emph{undirected} (e.g., protein-protein interaction networks). 
Thus, these approaches cannot be directly applied to such networks.

Analogous to extending the notion of network motifs from the static to
dynamic setting, recently, we took the first step to do the same with
the notion of graphlets. Namely, we used graphlets, along with several
other measures of network topology, as a basis of a static-temporal
approach to study human aging from protein-protein interaction
networks
\cite{Faisal2014dynamic}. That is, we counted static graphlets, 
along with the other network measures, within each snapshot (where
different snapshots correspond to different human ages), and then we
studied the time-series of the results to gain insights into network
structural changes with age \cite{Faisal2014dynamic}.  However, in
this initial work, we only used the already established notion of
static graphlets within a static-temporal approach that ignored
important relationships between different snapshots, in order to
demonstrate that accounting for at least some temporal information in
the static-temporal fashion can improve results compared to using the
trivial static (aggregate) approach that has traditionally been used
in the field of computational biology. Further important temporal
inter-snapshot information remains to be explored via a novel truly
temporal approach. In this study, we aim to develop such an approach,
as follows.

\subsection{Our contribution}

To overcome the issues of the existing methods for temporal network
analysis, we take the well established notion of static graphlets to
the next level to develop new theory of \textit{dynamic graphlets}
that are needed for efficient truly temporal network analysis.  Unlike
any of the existing temporal motif-based approaches, our dynamic
graphlets allow for
\emph{all} of the following: \textbf{1)} they can study topological and
temporal structures of \emph{arbitrary} complexity, as permitted by
available computational resources; \textbf{ 2)} there are \emph{no
limitations} such as the one on the number of events that a node can
participate in; \textbf{3)} they can capture the topological signature
of the entire network \emph{as well as} of each individual node;
\textbf{4)} they allow for studying temporal networks \emph{without}
relying on a null model; \textbf{5)} they complement the temporal
motif-based approaches by working with
\emph{undirected} networks. Unlike the existing graphlet-based 
static-temporal approach, our dynamic graphlets explicitly consider
relationships between different snapshots.

Compared to the existing methods, the closest approaches to our work
are those of temporal motifs as defined in \cite{kovanen2011temporal},
static graphlets \cite{prvzulj2004modeling,milenkovic2008uncovering},
and static-temporal graphlets
\cite{Faisal2014dynamic}.  Since temporal motifs have limitations (see above), 
the major one being dependence on a null model, they are not
comparable to our dynamic graphlet approach.  Static and
static-temporal graphlet approaches, which ignore all temporal
information or inter-snapshot temporal information, respectively, are
directly comparable to our dynamic graphlet approach, which does
account for inter-snapshot temporal information. Thus, our goal is to
fairly compare the three approaches, in order to evaluate the effect
on result accuracy of the amount of temporal information that the
given approach can consider.


In the rest of the paper, we formally define our novel notion of
dynamic graphlets and present an approach for enumerating all dynamic
graphlets of an arbitrary size and counting them in a given network
(Section~\ref{sec:methods}).  We thoroughly evaluate the ability of
dynamic graphlets to characterize the structure and function of an
entire temporal network as well as of individual nodes. Namely, on
both synthetic and real-world temporal network data, we measure how
well our approach can group (or cluster) temporal networks (or nodes)
of similar structure and function and separate those networks (or
nodes) of dissimilar structure and function. We find that our dynamic
graphlet approach outperforms both static and static-temporal graphlet
approaches in all of these tasks (Section~\ref{sec:results}). This
confirms our hypothesis that accounting for more temporal information
leads to better result accuracy. This in turn illustrates real-life
relevance of our new dynamic graphlet methodology, especially because
the amount of available temporal network data is expected to continue
to grow across many domains.

\section{Methods}\label{sec:methods}

We introduce the theoretic notion of dynamic graphlets in
Section~\ref{sec:methods_d_graphlets}.  We give an algorithm for
dynamic graphlet counting in
Section~\ref{sec:methods_counting_graphlets}.  We discuss a related
notion of \emph{causal} dynamic graphlets in
Section~\ref{sec:methods_dc_graphlets}.  We describe our experimental
setup and evaluation framework in
Section~\ref{sec:methods_experiment}.

\subsection{Dynamic graphlets}\label{sec:methods_d_graphlets}

Let $G(V,E)$ be a \emph{temporal network}, where $V$ is the set of
nodes and $E$ is the set of \emph{events} (temporal edges) that
are associated with a start time and duration
\cite{holme2012temporal}.  An event can be represented as a $4$-tuple
$(u,v,t_{start},\sigma)$,
where $u$ and $v$ are its endpoints, $t_{start}$ is its starting time,
and $\sigma$ is its duration.  Thus, each event is linked to a unique
edge in the aggregate static network, whereas each static edge may be
linked to multiple events.
Note that here we consider undirected events, but most ideas can be extended to directed events as well.

Let $G'(V',E')$ be a \emph{temporal subgraph} of $G$ with $V'
\subseteq V$ and $E' \subseteq E$, where $E'$ is restricted to nodes
in $V'$.  Let events $e_i$ and $e_j$ be \emph{$\Delta t$-adjacent} if
they share a node and if both events occur within a given time
interval $\Delta t$.  Two events $e_i$ and $e_j$ are $\Delta
t$-connected if there exists a sequence of $\Delta t$-adjacent events
joining $e_i$ an $e_j$.  A temporal network is called \emph{$\Delta
t$-connected} if any two of its nodes are $\Delta t$-connected.

Let two nodes $s$ and $t$ be connected by a \emph{$\Delta
t$-time-respecting path} if there is a sequence of events
$(v_0,u_0,t_{start0},\sigma_0),(v_1,u_1,t_{start1},\sigma_1),\dots,
(v_k,u_k,t_{startk},\sigma_k)$, such that $v_0=s$, $u_k=t$, $\forall i
\in [0,k-1]$ $u_i=v_{i+1}$, and $t_{i+1} \in
[t_{i}+\sigma_i,t_{i}+\sigma_i+\Delta t]$.  A temporal subgraph is
\emph{$\Delta t$-causal} if it has no isolated nodes and if for every
two events in this subgraph there exists a $\Delta t$-time-respecting
path containing both of the events.  So, every $\Delta t$-causal
subgraph is also $\Delta t$-connected, while the opposite is not
always true. 

\begin{figure}
  \centering\includegraphics[width=0.92\linewidth]{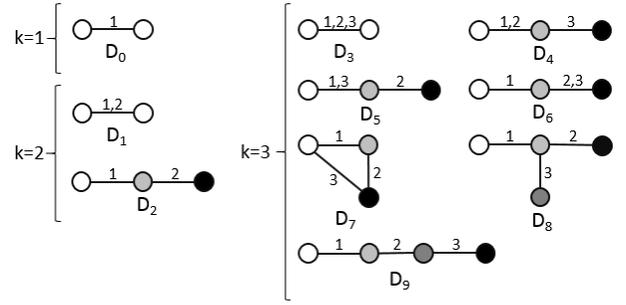}
  \caption{All dynamic graphlets with up to three events, along with their
  automorpism orbits.  Multiple events along the same edge are
  separated with commas.  Node colors correspond to different
  orbits. }\label{fig:dynamic_graphlets}
\end{figure}

Then, a \emph{dynamic graphlet} is an equivalence class of isomorphic
$\Delta t$-causal temporal subgraphs; equivalence is taken with
respect to the relative temporal order of events.  For isomorphism, we
do not consider events' actual start times but only their relative
ordering.  Thus, two $\Delta t$-causal temporal subgraphs will
correspond to the same dynamic graphlet if they are isomorphic and
their corresponding events occur in the same order.  Note that we
consider only $\Delta t$-causal temporal subgraphs, in contrast to
temporal motifs that consider only $\Delta t$-connected subgraphs
\cite{kovanen2011temporal}, so our definition is more restricting.
\figurename~\ref{fig:dynamic_graphlets} illustrates all dynamic graphlets with up to three events, but  we evaluate larger graphlets as well.

\begin{figure}
  \centering\includegraphics[width=0.92\linewidth]{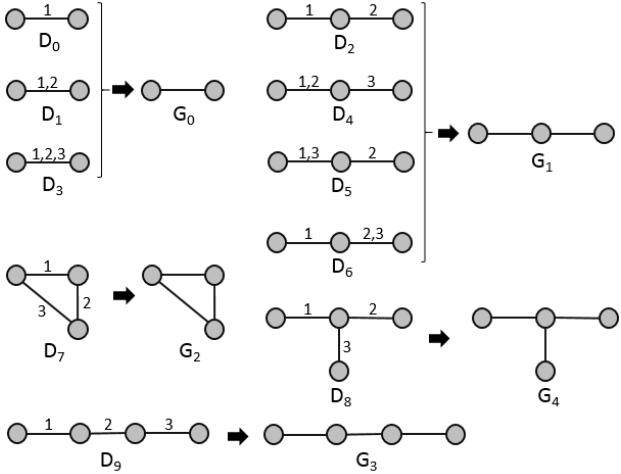}
  \caption{All dynamic graphlets $D_i$ with up to three events, and their
  corresponding (static) backbones $G_j$.  }\label{fig:backbones}
\end{figure}

Note that if for a given dynamic graphlet with $n$ nodes and $k$
events we discard the order of the events and remove duplicate events
over the same edge, we get a static graphlet with $n$ nodes and $k'
\leq k$ edges, which we call the \emph{backbone} of the dynamic
graphlet.  Each dynamic graphlet has a single backbone, while one
backbone can correspond to different dynamic graphlets
(\figurename~\ref{fig:backbones}).  Supplementary Table S1 shows for
each static graphlet with up to five nodes
\cite{milenkovic2008uncovering} the number of corresponding dynamic
graphlets with up to ten events.

The above definitions allow us to describe all dynamic graphlets of a
given size in the \emph{entire network}, in order to obtain
topological signature of the network. There already exists a popular
notion of topological signature of an \emph{individual node} in a
static network, called the graphlet degree vector (GDV) the node,
which describes the number of each of the static graphlets that the
node ``touches'' at a specific ``node symmetry group'' (or
automorphism orbit) within the given graphlet
(\figurename~\ref{fig:static_graphlets})
\cite{milenkovic2008uncovering}.  Analogously, one might want to
describe the node's \emph{dynamic} GDV equivalent. In this case,
automorphism orbits of a dynamic graphlet will be determined based on
both topological (as in static case) and temporal (unlike in static
case) position of a node within the dynamic graphlet.  Thus, a dynamic
graphlet with $n>2$ nodes will have $n$ different orbits
(\figurename~\ref{fig:dynamic_graphlets}), whereas the number of orbits in a
static graphlet of size $n$ is typically less than $n$
(\figurename~\ref{fig:static_graphlets}); note that for dynamic graphlets
with $n=2$, there will be only one orbit, since events are undirected
and thus their end nodes are topologically equivalent.

\begin{figure}
  \centering\includegraphics[width=\linewidth]{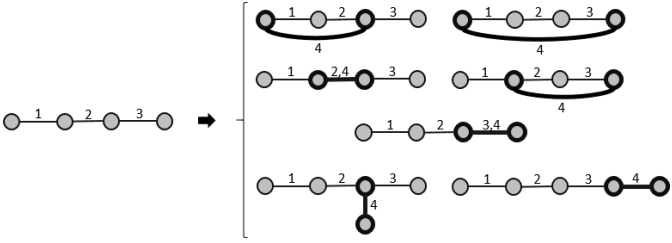}
  \caption{Illustration of how we extend a dynamic graphlet with an
  additional more recent event, on the example of $D_9$.  There are
  seven possible extensions of $D_9$ (which contains four nodes and
  three events) with the most recent event $4$ (shown in bold) into a
  dynamic graphlet with four events. Five of the extensions keep the
  same number of nodes but increment the number of events, while the
  remaining two extensions increment both the number of nodes and
  events.  Note that in order to extend $D_9$ with event $4$, at least
  one of the nodes involved in event $3$ has to participate in event
  $4$ as well.  }\label{fig:extending_graphlet}
\end{figure}

Next, we discuss the way for computing $D(n,k)$, the number of dynamic
graphlets with $n$ nodes and $k$ events.  Since at least $n-1$ edges
are needed to connect $n$ nodes, it follows that $D(n,k)=0$ for
$k<n-1$.  Moreover, since we assume that events are undirected,
$D(2,k)=1$, for any $k$.  To compute $D(n,k)$ when $n \geq 3$ and $k
\geq n-1$, notice that each dynamic graphlet with $k$ events can be
formed from a dynamic graphlet with $k-1$ events and either $n-1$ or
$n$ nodes, by adding a new event between some two existing nodes or
between an existing node and a new node, respectively
(\figurename~\ref{fig:extending_graphlet}).
  
In the first case, we take a dynamic graphlet with $n$ nodes and $k-1$
events and add a new event between its existing nodes, in order to
obtain a dynamic graphlet with $n$ nodes and $k$ events (e.g.,
construct $D_6$ from $D_2$).  Due to the $\Delta t$-causality
constraint, this new event has to involve at least one of the two
nodes participating in event $(k-1)$.  We can add the new event in
$2n-3$ different ways: between one of these nodes and the
``remaining'' $n-2$ nodes (which is $2(n-2)=2n-4$ ways) or just
duplicate event $(k-1)$.
  
In the second case, we take a dynamic graphlet with $n-1$ nodes and
$k-1$ events and add an event from one of its nodes to the new
($n^{th}$) node, in order to obtain a dynamic graphlet with $n$
nodes and $k$ events (e.g., construct $D_4$ from $D_1$).  For $n-1 \geq 3$,
there are two ways to do this, since there are two potential
candidates for this new event (the two endpoints of event
$(k-1)$). Note that since events are undirected, for $n-1=2$, the two
nodes are indistinguishable from our point of view, and so we have
only one way to construct a new dynamic graphlet with $3$ nodes and
$k$ events.

In summary, we can get $2n-3$
new dynamic graphlets with $n$ nodes and $k$ events from each dynamic
graphlet with $n$ nodes and $k-1$ events.
Moreover, we can get two new dynamic graphlets with $n$ nodes and
$k$ events from each dynamic graphlet with $n-1$ nodes and $k-1$
events.  The only exception is for $n=3$, since we can get only one
new dynamic graphlet with three nodes from a dynamic graphlet with two
nodes, as these two nodes are indistinguishable (\figurename~\ref{fig:dynamic_graphlets}).
Importantly, since each dynamic graphlet with $k$ events has a unique $(k-1)$-``prefix'' from
which it was extended, all of these new dynamic graphlets with $n$
nodes will be different.  Thus, we get the following recursive
formulas for $D(n,k)$:

\vspace{-0.3cm}

$$ D(3,k) = 3D(3,k-1) + D(2,k-1), n=3$$

\vspace{-0.6cm}

$$ D(n,k) = (2n-3)D(n,k-1) + 2D(n-1,k-1), n>3$$

By expanding the formulas for few smallest values of $n$ and $k$, we
can get the following closed-form solution:

\vspace{-0.3cm}

$$ D(n,k) = \sum\limits_{i=0}^{n-2} \frac{(-1)^{n+i} \binom{n-2}{i}
(2i+1)^{k-1}}{2(n-2)!} , n \geq 3. $$

Supplementary Table S2 shows dynamic graphlet counts for up to $n=11$
and $k=10$.

Since now we can compute $D(n,k)$, we next consider the task of
enumerating and generating each of these dynamic graphlets (we discuss
the process of counting each of the generated graphlets in Section
\ref{sec:methods_counting_graphlets}).  We build upon the fact that each
dynamic graphlet with $k$ events has a unique $(k-1)$-``prefix'' (see
above).  Thus, we start with a single event (dynamic graphlet $D_0$
with $n=2$ and $k=1$) as the current graphlet and then recursively
extend the current graphlet until the desired size is reached.  
Supplementary Algorithm S1 illustrates our enumeration
procedure.


\subsection{Counting dynamic graphlets in a network}\label{sec:methods_counting_graphlets}

As now we know the number of different dynamic graphlets with a given
number of nodes and events and also how to enumerate and generate each
one of them, how to actually count each of the dynamic graphlets in a
given network?

\begin{figure}
  \centering\includegraphics[width=\linewidth]{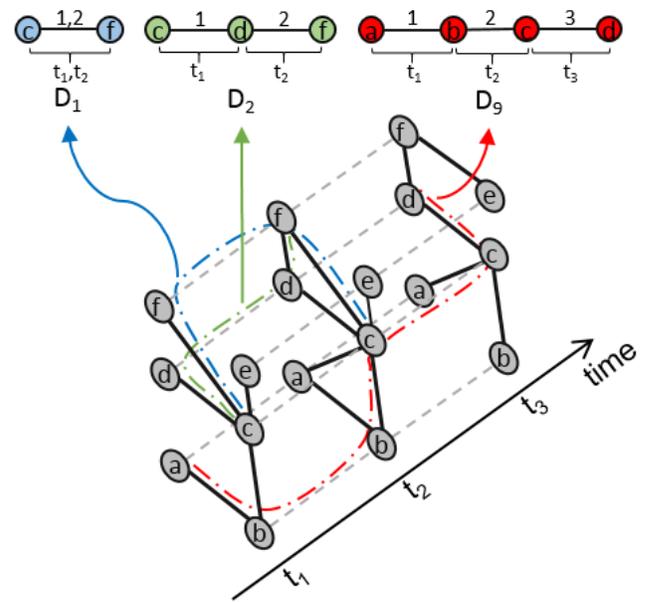}
  \caption{Illustration of our dynamic graphlet counting procedure.
    The temporal network is presented as a sequence of three
  snapshots.  Dashed lines denote instances of the same node in
  different snapshots.  Colored lines illustrate the path of how the
  temporal network is explored in order to count the given dynamic
  graphlet.  Regular dynamic graphlet counting (Section~\ref{sec:methods_counting_graphlets}) will detect all three
  of the dynamic graphlets $D_1$ (involving nodes $c$ and $f$), $D_2$
  (involving nodes $c$, $d$, and $f$), and $D_9$ (involving nodes $a$,
  $b$, $c$, and $d$).  Causal dynamic graphlet counting (Section~\ref{sec:methods_dc_graphlets}) will detect
  only the first two of these dynamic graphlets. This is because nodes
  $c$ and $d$ are interacting in both the second and third snapshot,
  and thus, the third condition from the definition of causal dynamic
  graphlet counting is violated when it comes to
  $D_9$.}\label{fig:graphlet_counting}
\end{figure}

We perform dynamic graphlet counting in the same way as we generate
the graphlets.  That is, for each event in a temporal network, we use
this event as the current dynamic graphlet $D_0$ and then search for
larger graphlets that are grown recursively from the current one
(\figurename~\ref{fig:graphlet_counting}).  Supplementary Algorithms
S2-S4 describe this algorithm. Its running time depends on the
structure of the given temporal network.  In general, since the
algorithm explicitly goes through every dynamic graphlet that it
counts, the running time is proportional to the  number of
dynamic graphlets.  For a network with $D$ $\Delta t$-adjacent event
pairs, counting all dynamic graphlets with up to $k$ events takes
$\mathcal{O}(|E| + |E|(\frac{D}{|E|} )^{k-1})$.  As with static
graphlets, the running time of exhaustive dynamic graphlet counting is
exponential in graphlet size (but is still practical, as we will
show).  Yet, as elegant non-exhaustive approaches were proposed for
faster static graphlet counting
\cite{hovcevar2014combinatorial, marcus2012rage}, similar techniques
can be sought for dynamic graphlet counting as well.

\subsection{Causal counting of dynamic graphlets in a network}\label{sec:methods_dc_graphlets}
 A network having dense neighborhoods with many events
between the same node pairs will have a large number of different
dynamic graphlets with large counts.  This is because for a given
dynamic graphlet, there will be many $\Delta t$-adjacent candidates
which can be used to ``grow'' this dynamic graphlet.  For each of
these possible extensions of a dynamic graphlet, we will again have
many possibilities for further extension, and so on.  For example,
consider a snapshot-based network representation where each snapshot
is the same dense graph.  Clearly, a large number of different dynamic
graphlets will be detected, yet many of them will just be artifacts of
the consecutive snapshots ``sharing'' the dense network structure.

To address this issue and remove the likely redundant graphlet counts,
which is expected to also reduce computational complexity of dynamic
graphlet counting, we propose a modification to the counting process
from Section~\ref{sec:methods_counting_graphlets}, as follows. When we
are extending a dynamic graphlet ending with event
$e_1=(u_1,v_1,t_1,\sigma_1)$ with a new event
$e_2=(u_2,v_2,t_2,\sigma_2)$, if $\{u_1,v_1\} =
\{u_2,v_2\}$, i.e., if the two events correspond to the same static edge, 
then we impose the same two conditions as in the regular counting
procedure from Section \ref{sec:methods_counting_graphlets}: 1) the
two events $e_1$ and $e_2$ must be $\Delta t$-adjacent with $t_2>t_1$,
and 2) the two events must share a node. Otherwise, if $\{u_1,v_1\}
\neq
\{u_2,v_2\}$, i.e., if the two events correspond to two different static 
edges, we also add a new third condition: to extend the dynamic
graphlet ending with event $e_1$ with event $e_2$, $u_2$ and $v_2$
cannot interact between the starting times of $e_1$ and $e_2$ (i.e.,
$\not\exists e'=(u_2,v_2,t',\sigma')
\in E$ with $ t_1 -
\sigma' \leq t' < t_2$).

Intuitively, the new third condition requires a ``causal''
relationship between $e_1$ and $e_2$: $u_2$ and $v_2$ start their
interaction only after the end of $e_1$ (though note that there could
still be an event involving $u_2$ and $v_2$ sometime before the start
of $e_1$).  That is, in order to extend a dynamic graphlet ending with
event $e_1$ with some event $e_2$, the two nodes participating in
$e_2$ should not interact with each other between the start of $e_1$
and the start of $e_2$, unless $e_1$ and $e_2$ involve the same nodes
(otherwise, the counting process is as in Section
\ref{sec:methods_counting_graphlets}). This allows one to reduce the
number of likely redundant temporal subgraphs that are being
considered, which in turn reduced the total running time.

Note that we split the counting procedure into two above cases ($e_1$
and $e_2$ corresponding to the same static edge, and $e_1$ and $e_2$
not corresponding to the same edge) for the following reason.  We want
to impose the new third condition only in the latter case, but not in
the former one. This is because in the former case we still want to
allow for counting dynamic graphlets having consecutive repetitions of
the same event, such as $D_1$ or $D_4$. And if we imposed the third
condition in the former case as well, then such a dynamic graphlet
would never be counted.

Henceforth, we refer to this modified counting procedure as
\textit{causal dynamic graphlet counting}.
\figurename~\ref{fig:graphlet_counting} illustrates the distinction between regular and causal dynamic graphlet counting procedures. Clearly, causal dynamic graphlet counting 
allows for examining fewer dynamic graphlet options during counting
compared to regular dynamic graphlet counting, because the former
excludes from consideration graphlets that are likely artifacts of
repeated events, unlike the latter. As a consequence, causal dynamic
graphlet counting is expected to be more computationally efficient in
terms of running time.

\subsection{Experimental setup}\label{sec:methods_experiment}

\noindent\textbf{Graphlet methods under consideration, 
and the corresponding network construction strategies.}  We compare
four graphlet-based approaches: static, static-temporal, dynamic and
causal dynamic graphlets.  To apply static graphlet counting to a
temporal network, we first aggregate the temporal data into a single
static network, by keeping the node set the same, and by adding an
edge between two nodes in the static network if there are at least $w$
events between these two nodes in the temporal network.  For other
methods, we use a snapshot-based representation of the temporal
network: we split the whole time interval of the temporal network into
time windows of size $t_w$, and for each time window, we construct the
corresponding static snapshot by aggregating the temporal data during
this window with the parameter $w$, as above.  For static and
static-temporal graphlet approaches, we vary the number of graphlet
nodes $n$, and for dynamic and causal dynamic graphlet approaches, we
vary both the number of graphlet nodes $n$ and the number of graphlet
events $k$.

We began our analysis by testing in detail $w$ values of 1, 2, 3, 5,
and 10 on one of our data sets (see below). Since we observed no
qualitative differences in results produced by the different choices
of this parameter, we continued with the choice of $w=1$, and we
report the corresponding results throughout the paper. We also tested
multiple values for $t_w$ in each data set (see below), and again we
saw no significant qualitative differences in the results. Hence,
throughout the paper, we report results for $t_w=2$ (the unit of time
for this parameter depends on the data set; see below).

\noindent\textbf{Network classification.} An approach that captures 
network structure (and function) well should be able to group together
similar networks (i.e., networks from the same class) and separate
dissimilar networks (i.e., networks from different classes)
\cite{yaverouglu2014revealing}. To evaluate our dynamic graphlet
approach against static and static-temporal graphlet approaches in
this context, we generate a set of synthetic (random graph) temporal
networks of nine different classes corresponding to nine different
versions of an established network evolution model
\cite{leskovec2008microscopic}. We use synthetic temporal network data 
because obtaining real-world temporal network data for multiple
different classes and with multiple examples per class is hard. And
even if a wealth of temporal network data were available, we typically
have no prior knowledge of which real-networks are (dis)similar, i.e.,
which networks belong to which functional class.

The network evolution model that we use was designed to simulate
evolution of real-world (social) networks, and it incorporates the
following parameters: node arrival rate, initiation of an edge by a
node, and selection of edge destination. Specifically, the model is
parameterized by the node arrival function $N(t)$ that corresponds to
the number of nodes in the network at a given time, parameter
$\lambda$ that controls the lifetime of a node, and parameters
$\alpha$ and $\beta$ that control how active the nodes are in adding
new edges. By choosing different options for the model parameters, we
can generate networks with different evolution processes.  In
particular, for our analysis, we test three different types of the
node arrival function (linear, quadratic, and exponential) and two
sets of parameters corresponding to edge initiation
($\lambda_1=0.032$, $\alpha_1=0.8$, $\beta_1=0.002$, and
$\lambda_2=0.02$, $\alpha_2=0.9$, $\beta_2=0.004$)
\cite{leskovec2008microscopic}, resulting in six different network
classes.  We also test a modification of the network evolution model,
in which each node upon arrival simply adds a fixed number of edges
(in our case, 20) according to preferential attachment and then stops \cite{barabasi1999emergence}.
Intuitively, this modification corresponds to preferential attachment
model extended with a node arrival function.  In this way, we create
three additional network classes, one for each of the three node
arrival functions, resulting in nine different network classes in
total.  

In order to test the robustness of the network classification methods
to the network size, in each of the nine classes, we test three
network sizes: 1000, 2000, and 3000 nodes.  Then, for each network
size and class, we generate 25 random graph instances.  For the above
synthetic network set, we report results for the following network
construction parameters: $w=1$ and $t_w=2$.  We tested other parameter
values as well ($t_w=5$ and $t_w=10$), and all results were
qualitatively similar. Also, unless otherwise noted, we report results
for the largest network size of 3000 nodes. Results for the other
network sizes were qualitatively similar.

Given the resulting aggregate or snapshot-based network
representations, we then compute static, static-temporal, or dynamic
graphlet counts in each network and reduce the dimensionality of the
networks' graphlet vectors with principle component analysis
(PCA). For a given graphlet vector, we keep its first two PCA
components, since in all cases the first two PCA components account
for more than 90\% of variation.  Then, we use Euclidean distance in
this PCA space as a network distance measure and evaluate whether
networks from the same class are closer in the graphlet-based PCA
space than networks from different classes, as described below.

\noindent\textbf{Node classification.}
 We also compare the three graphlet-based methods by evaluating
 whether they can group together similar \emph{nodes} rather than
 entire networks. Specifically, we measure the ability of the methods
 to distinguish between functional node labels (i.e., classes) based
 on the nodes' graphlet-based topological signatures. As a proof of
 concept, we do this on a publicly available Enron dataset
 \cite{priebe2005scan}, which is both temporal and contains node
 labels. Unfortunately, availability of additional temporal and
 labeled network data is very limited. The Enron network is based on
 email communications of 184 users from 2000 to 2002, and seven
 different user roles in the company are used as their labels: CEO,
 president, vice president, director, managing director, manager, and
 employee.

For the above real-world network, we report results for the following
network construction parameters: $w=1$ and $t_w=2$ months.  Note that
we tested other parameter values as well ($w=2$, $w=3$, $w=5$, and
$w=10$; $t_w=1$ week, $t_w=2$ weeks, $t_w=1$ month, and $t_w=3$
months), and all results were qualitatively similar.

Given the appropriate aggregate or snapshot-based network data
representations, we then compute static, static-temporal, or dynamic
graphlet counts of each node in the network and reduce the
dimensionality of the given node's graphlet vector with PCA. We keep
the first three PCA components to account for enough of the variation.
Then, we use Euclidean distance in this PCA space as a node distance
measure, and evaluate whether nodes having the same label are closer
in the graphlet-based PCA space than nodes with different labels, as
follows.

\noindent\textbf{Evaluation strategy.}
We have a set of objects (networks or nodes), graphlet-based PCA
distances between the objects, and the objects' ground truth
classification (with respect to nine network classes or seven node
labels). For a given method, we measure its graphlet-based PCA
performance as follows.

First, we take all possible \emph{pairs} of objects and retrieve them
in the order of increasing distance, starting from the closest
ones. We retrieve the object pairs in increments of $k\%$ (including
ties), where we vary $k$ from $0\%$ to $100\%$ in increments of 0.01\%
until we retrieve top 1\% of all pairs and in increments of 1\%
afterwards. If we retrieve a pair with two objects of the same ground
truth class, the pair is a true positive, otherwise the pair is a
false positive. At a given step, for all pairs that we do not
retrieve, the given pair is either a true negative (if it contains
objects of different classes) or a false negative (if it contain
objects of the same class).  Then, at each value of $k$, we compute
precision, the fraction of correctly retrieved pairs out of all
retrieved pairs, and recall, the fraction of correctly retrieved pairs
out of all correct pairs. We find the value of $k$ where precision and
recall are equal, and we refer to the resulting precision and recall
value as the break-even point. Since lower precision means higher
recall, and vice versa, we summarize the two measures into F-score,
their harmonic mean, and we report the maximum F-score over all values
of $k$.  To summarize these results over the whole range of $k$, we
measure average method accuracy by computing the area under the
precision-recall curve (AUPR).  Moreover, we compute an alternative
classification accuracy measure, namely the area under the receiver
operator characteristic curve (AUROC), which corresponds to the
probability of a method ranking a randomly chosen positive pair higher
than a randomly chosen negative pair (and so the AUROC value of $0.5$
corresponds to a random result). AUPRs are considered to be more
credible than AUROCs when there exists imbalance between the size of
the set of network pairs that share a class and the size of the set of
network pairs that do not share a class.

Second, we split all pairs of objects (and their corresponding
graphlet-based PCA distances) into two classes: correct pairs (each
containing two objects of the same class) and incorrect pairs (each
containing two objects of two different classes).  Then, we compare
distances of correct pairs against distances of incorrect pairs, with
the expectation that distances of the correct pairs would be
statistically significantly lower than distances of the incorrect
pairs. For this purpose, we compare the two sets of distances using
Wilcoxon rank-sum test \cite{hulovatyy2014network}.

For each of these evaluation tests, we also evaluate all three
graphlet-based methods against a random approach.  First, as the
simplest possible random approach (which favors the graphlet-based
methods the most), we randomly embed objects (networks or nodes) into
a 2-dimensional (for networks) or 3-dimensional (for nodes) Euclidean
space, compute the objects' pairwise Euclidean distances, and evaluate
the resulting random approach in the same way as above.  Second, as a
more sophisticated and restrictive random approach (which favors the
graphlet-based methods the least), for each graphlet-based method, we
keep its actual PCA distances between objects, and we just randomly
permute the object classes/labels before we evaluate the results.  By
comparing the performance of each actual method with the performance
of the method's corresponding restrictive random counterpart, we can
be more confident in potential non-random behavior of the given method
than with the initial simple randomization approach.  

For each randomization approach, we compute its results as an average
over 10 different runs.  We report as ``random'' approach's results
the highest-scoring values over all of the different randomization
schemes, in order to gain as much confidence as possible into the
graphlet approaches' results.

\section{Results and discussion}\label{sec:results}

We evaluate our novel dynamic graphlet approach against the existing
static and static-temporal graphlet approaches in the context of two
evaluation tasks: network classification
(Section~\ref{sec:results_network_classification}) and node
classification (Section~\ref{sec:results_node_classification}). Also,
we discuss the effect of different method parameters on the results
(Section~\ref{sec:results_effects}).

\subsection{Network classification}\label{sec:results_network_classification}

\begin{figure}
  \centering\includegraphics[width=0.7\linewidth,
  angle=270]{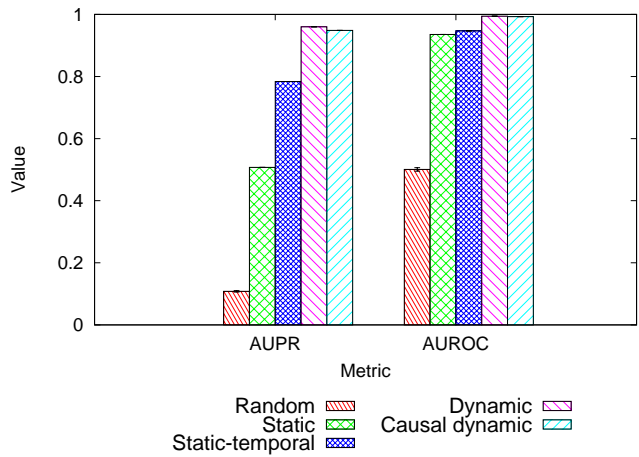} \caption{Network classification
  accuracy of the different methods in terms of AUPR and AUROC. For
  each method, the highest-scoring graphlet size is chosen. For other
  parameter choices, see Table \ref{tab:model_analysis_results}.}
  \label{fig:model_aupr_auroc}
\end{figure}

\begin{figure}
  \centering\includegraphics[width=0.7\linewidth,
  angle=270]{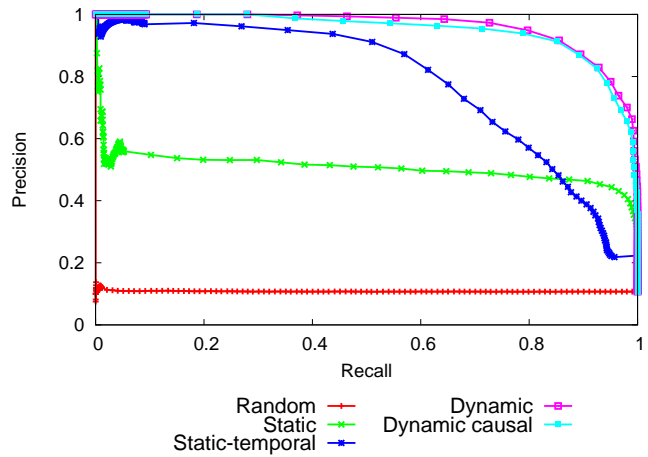} \caption{Network classification accuracy of
  the different methods in terms of precision-recall curves. For each
  method, the highest-scoring graphlet size is chosen. For other
  parameter choices, see Table \ref{tab:model_analysis_results}. }
  \label{fig:model_3na_3wa_3000_p_r}
\end{figure}

We test how well the different methods distinguish between nine
different classes of synthetic temporal networks based on the
networks' graphlet counts. The different evaluation criteria give
consistent results: while according to Wilcoxon rank-sum test, all
methods have intra-class distances significantly lower than
inter-class distances and thus show non-random behavior ($p$-values
less than $10^{-100}$), (causal) dynamic graphlets are superior both
in terms of accuracy and computational complexity, followed by
static-temporal graphlets, followed by static graphlets,
(\figurename~\ref{fig:model_aupr_auroc},
\figurename~\ref{fig:model_3na_3wa_3000_p_r}, and
Table~\ref{tab:model_analysis_results}). Some additional observations
are as follows: regular dynamic graphlets perform better than causal
dynamic graphlets in terms of accuracy, and the two are comparable in
terms of computational complexity. 

\begin{table*}
\centering
\begin{tabular}{l||cccccc}
  Method & AUPR & AUROC & Break-even point & Maximum F-score & Running time, s     \\
  \hline
  \hline
  Static, 3-node & 0.507 & 0.935 & 0.508 & 0.613 & 3.3 (1.785)  \\ 
  Static, 4-node & 0.423 & 0.882 & 0.463 & 0.468 & 3.3 (1.785)   \\
  Static, 5-node & 0.321 & 0.807 & 0.341 & 0.376 & 3.3 (1.785)   \\
  \hline
  Static-temporal, 3-node & 0.784 & 0.947 & 0.702 & 0.707 & 3.7 (0.173)  \\
  Static-temporal, 4-node & 0.498 & 0.826 & 0.475 & 0.476 & 3.7 (0.173)   \\
  Static-temporal, 5-node & 0.374 & 0.790 & 0.379 & 0.390 & 3.7 (0.173)   \\
  \hline
  Dynamic, 3-event, 3-node & \textbf{0.960} & \textbf{0.994} & \textbf{0.884} & \textbf{0.885} & \textbf{0.6 (0.116)}  \\
  Dynamic, 5-event, 3-node & \textbf{0.960} & \textbf{0.994} & \textbf{0.884} & \textbf{0.885} & 0.7 (0.104) \\
  Dynamic, 7-event, 3-node & \textbf{0.960} & \textbf{0.994} & \textbf{0.884} & \textbf{0.885} & 1.4 (0.149) \\
  Dynamic, 6-event, 4-node & 0.714 & 0.937 & 0.656 & 0.660 & 4.8 (0.875)  \\
  \hline
  Causal dynamic, 3-event, 3-node & 0.949 & 0.993 & 0.881 & 0.881 & \textbf{0.6 (0.188)}  \\
  Causal dynamic, 5-event, 3-node & 0.949 & 0.993 & 0.881 & 0.881 & 0.7 (0.145)  \\
  Causal dynamic, 7-event, 3-node & 0.949 & 0.993 & 0.881 & 0.881 & 1.5 (0.206) \\
  Causal dynamic, 6-event, 4-node & 0.740 & 0.939 & 0.672 & 0.675 & 4.2 (0.684)  \\
  \hline
  Random & 0.107 (0.002) & 0.499 (0.005) & 0.108 (0.006) & 0.194 (0.000) & -  \\
\end{tabular}

  \caption{Detailed network classification results for the different
  methods and different parameters in each method. Different columns
  correspond to different performance measures. In a given column, the
  value in bold corresponds to the best result over all methods.
  Numbers in parentheses correspond to standard deviations. For
  illustration purposes, graphlet counting running times are shown for
  one of the nine network classes (using the exponential node addition
  function and the first set of edge initiation parameters
  (Section~\ref{sec:methods_experiment})); running times for the
  remaining network classes are shown in Supplementary Table S3.  Note
  that for static and static-temporal graphlets, running times for 3-
  and 4-node graphlets are the same as for 5-node graphlets simply
  because their implementations \cite{GraphCrunch,GraphCrunch2}
  compute graphlet counts for all up to 5-node graphlets by default
  and then they compute graphlet counts for smaller graphlet sizes
  simply by removing counts corresponding to the larger graphlet
  sizes.  }
  \label{tab:model_analysis_results}
\end{table*}

\subsection{Node classification}\label{sec:results_node_classification}

\begin{figure}
  \centering\includegraphics[width=0.7\linewidth,
  angle=270]{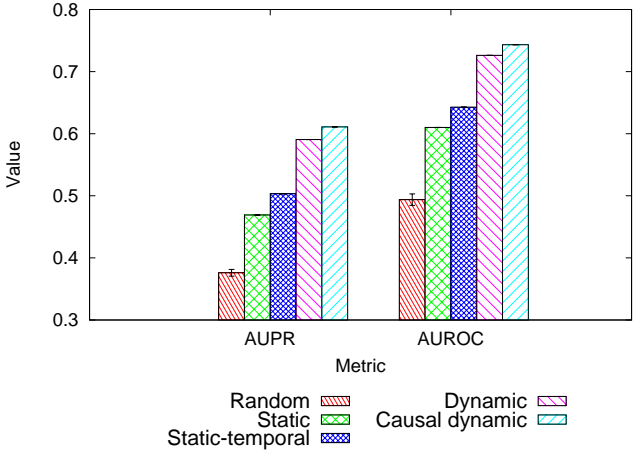} \caption{Node classification accuracy
  of the different methods in terms of AUPR and AUROC.  For each
  method, the highest-scoring graphlet size is chosen. For other
  parameter choices, see Table~\ref{tab:node_analysis_results}.  }
  \label{fig:enron_aupr_auroc}
\end{figure}

\begin{figure}
  \centering\includegraphics[width=0.7\linewidth,
  angle=270]{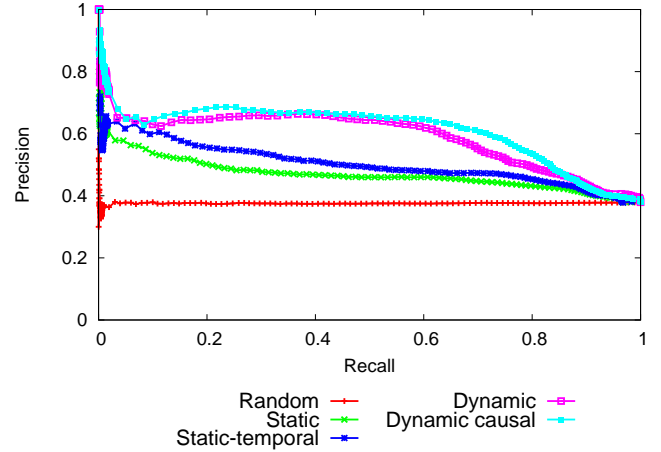} \caption{Node classification accuracy of the
  different methods in terms of precision-recall curves. For each
  method, the highest-scoring graphlet size is chosen. For other
  parameter choices, see Table~\ref{tab:node_analysis_results}. }
  \label{fig:enron_p_r}
\end{figure}

Also, we test how well the different methods distinguish between six
different classes of nodes in a real-world network based on the nodes'
graphlet counts. The different evaluation criteria give consistent
results: while according to Wilcoxon rank-sum test, all methods have
intra-class distances significantly lower than inter-class distances
and thus show non-random behavior ($p$-values less than $10^{-100}$),
just as with network classification, (causal) dynamic graphlets are
again superior both in terms of accuracy and computational complexity,
followed by static-temporal graphlets, followed by static graphlets
(\figurename~\ref{fig:enron_aupr_auroc}, \figurename~\ref{fig:enron_p_r}, and
Table~\ref{tab:node_analysis_results}).

\begin{table*}
\centering
\begin{tabular}{l||cccccc}
  Method & AUPR & AUROC & Break-even point & Maximum F-score & Running time, s   \\
  \hline
  \hline
  Static, 3-node & 0.464 & 0.600 & 0.456 & 0.562 & 9.4 \\ 
  Static, 4-node & 0.469 & 0.610 & 0.461 & 0.567 & 9.4  \\
  Static, 5-node & 0.464 & 0.604 & 0.462 & 0.566 & 9.4  \\
  \hline
  Static-temporal, 3-node & 0.499 & 0.644 & 0.508 & 0.571 & 2.7  \\
  Static-temporal, 4-node & 0.503 & 0.643 & 0.609 & 0.689 & 2.7  \\
  Static-temporal, 5-node & 0.482 & 0.570 & 0.486 & 0.554 & 2.7  \\
  \hline
  Dynamic, 3-event, 3-node & 0.479 & 0.622 & 0.477 & 0.569 & 2.7  \\
  Dynamic, 5-event, 3-node & 0.474 & 0.615 & 0.458 & 0.569 & 9.6  \\
  Dynamic, 7-event, 3-node & 0.470 & 0.609 & 0.460 & 0.572 & 27.5 \\
  Dynamic, 3-event, 4-node & 0.541 & 0.684 & 0.547 & 0.594 & 24.5  \\  
  Dynamic, 6-event, 4-node & 0.525 & 0.666 & 0.516 & 0.583 & 1,024  \\
  Dynamic, 4-event, 5-node & 0.591 & 0.726 & 0.615 & 0.620 & 753  \\
  \hline
  Causal dynamic, 3-event, 3-node & 0.491 & 0.639 & 0.498 & 0.569  & \textbf{1.1} \\
  Causal dynamic, 5-event, 3-node & 0.492 & 0.638 & 0.495 & 0.570 & 1.9 \\
  Causal dynamic, 7-event, 3-node & 0.492 & 0.638 & 0.495 & 0.571 & 2.6 \\
  Causal dynamic, 3-event, 4-node & 0.550 & 0.695 & 0.570 & 0.600 & 4.9  \\
  Causal dynamic, 6-event, 4-node & 0.550 & 0.695 & 0.571 & 0.600 & 37.2 \\
  Causal dynamic, 4-event, 5-node & 0.594 & 0.732 & 0.618 & 0.637  & 60.8 \\
  \hdashline 
  Causal dynamic, 5-event, 6-node & \textbf{0.611} & \textbf{0.743} & \textbf{0.636} & \textbf{0.654} & 815  \\
  Causal dynamic, 6-event, 7-node & 0.608 & 0.742 & 0.635 & 0.652 & 10,029 \\
  \hline
  Random & 0.376 (0.009) & 0.495 (0.016) & 0.369 (0.007) & 0.550 (0.000) & -\\
\end{tabular}
  \caption{Detailed node classification results for the different
  methods and different parameters in each method. The table can be
  interpreted just as Table~\ref{tab:model_analysis_results}.  Causal
  dynamic graphlet methods below the dashed line correspond to
  parameter choices that were not feasible to test with regular
  dynamic graphlets. Also, notice that in this test of node
  classification we could test some additional parameters (e.g.,
  graphlets on five or more nodes) compared to the test of network
  classification (Table~\ref{tab:model_analysis_results}); this is
  because the test of network classification is computationally much
  more complex, given that graphlets need to be counted in multiple
  networks, as opposed to counting graphlets in only one network in
  the node classification task.}
  \label{tab:node_analysis_results}
\end{table*}

In this evaluation test of node classification, unlike in the test of
network classification, the best parameter version of causal dynamic
graphlets is more accurate than the best parameter version of dynamic
graphlets. We note, however, that due to the differences in the
counting process, causal dynamic graphlet counting allows us to
consider larger graphlet sizes (e.g., six or seven nodes) that are not
attainable when using regular dynamic graphlet counting due to computational constraints
(Table~\ref{tab:node_analysis_results}).  And it is at these large
graphlet sizes of six or seven nodes where causal dynamic graphlets
peform the best. So, in order to evaluate which one is more accurate,
dynamic graphlets or causal dynamic graphlets, it might not be fair to
compare the two methods' best parameter versions, due to differences
in the considered graphlet sizes.  Nonetheless, even if we compare
dynamic and causal dynamic graphlets of the same size, we find that
causal dynamic graphlets still demonstrate better results
(Table~\ref{tab:node_analysis_results}).

Further, in this evaluation test of node classification, unlike in the
test of network classification, causal dynamic graphlet counting takes
significantly less time than regular dynamic graphlet counting
(Table~\ref{tab:node_analysis_results}), which justifies our
motivation behind causal dynamic graphlets.

\begin{figure}
  \centering\includegraphics[width=\linewidth]{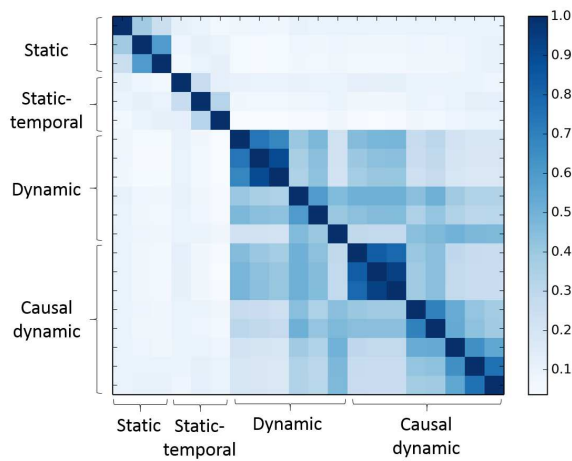}
  \caption{Pairwise similarities between the different methods and
  their parameter variations in the test of node
  classification. Similarities are computed as Jaccard similarity
  coefficients between two methods' top 5\% node pairs that are the
  closest in the graphlet-based PCA space.  The order of the methods
  in the figure directly corresponds to the method order in
  Table~\ref{tab:node_analysis_results} (we leave out detailed method
  names for visual clarity).  } \label{fig:enron_intersections}
\end{figure}

Importantly, not only the different methods differ quantitatively, but they also lead to different qualitative results (\figurename~\ref{fig:enron_intersections}): there is a clear separation between static, static-temporal, and (causal) dynamic graphlets in terms of which nodes they describe as topologically similar.
If we zoom into these results even further, within both
dynamic and causal dynamic graphlets, we can see two clear clusters
corresponding to three-node graphlets with different numbers of
events.  Thus, the number of nodes seems to play a larger role in
separating the different dynamic graphlet methods compared to the
number of events. (We discuss the effect of the method parameters in
more detail in the following section.)

\subsection{Effect of graphlet size on results}\label{sec:results_effects}

\noindent\textbf{Number of graphlet nodes.}
We next test the effect of graphlet size in terms of the number of
nodes on the result quality (i.e., accuracy), for all four graphlet
methods.  For network classification, we surprisingly find that
increasing the number of graphlet nodes leads to inferior accuracy,
for all three graphlet-based methods
(Table~\ref{tab:model_analysis_results}). On the other hand, in node
classification, for static and static-temporal graphlets, results are
almost the same for all graphlet sizes, with 4-node graphlets showing
marginally better performance, while for dynamic and causal dynamic
graphlets, larger number of nodes leads to better accuracy
(Table~\ref{tab:node_analysis_results}). 
In terms of the running time (rather than accuracy),
as expected, larger number of nodes leads to increase in computational
complexity, independent on evaluation test or graphlet method.

\noindent\textbf{Number of graphlet events.}
Also, we test the effect of graphlet size in terms of the number of
events on the result quality (i.e., accuracy), for dynamic and causal
dynamic graphlets (the other two graphlet methods, static and
static-temporal graphlets, do not deal with the notion of events,
i.e., temporal edges).  For network classification, we surprisingly
find that the number of events does not affect the accuracy
(Table~\ref{tab:model_analysis_results}).  On other hand, in node
classification, for a fixed number of nodes, the increase in the
number of events leads to slight improvement in accuracy for dynamic
graphlets but slight decrease in accuracy for causal dynamic graphlets
(Table~\ref{tab:node_analysis_results}). 
In terms of the running time (rather than accuracy), larger number of
events leads to increase in computational complexity, although the
level of running time increase is less pronounced than when increasing
the number of nodes.

\section{Conclusions}
The increasing availability of temporal real-world network data has
raised new challenges to the network researchers.  While one can use
the existing static approaches to study the aggregate or
snapshot-based network representation of the temporal data, doing so
overlooks important temporal information from the data.  Hence, we
develop a novel approach of dynamic graphlets that can capture the
temporal information  explicitly. In a systematic and thorough
evaluation, we demonstrate the superiority of our approach over its
static counterparts. This confirms that efficiently accounting for
temporal information helps with structural and functional
interpretation of the network data. This in turn illustrates real-life
relevance of our new dynamic graphlet methodology, especially because
the amount of available temporal network data is expected to continue
to grow across many domains.

\bibliographystyle{IEEEtran}
\bibliography{IEEEabrv,bibliography}

\end{document}